\documentclass[twocolumn]{aastex631}

\shorttitle{Orbits of Six Triple Systems}

\begin{document}

\renewcommand{\topfraction}{1.0}
\renewcommand{\bottomfraction}{1.0}
\renewcommand{\textfraction}{0.0}

\newcommand{\kms}{km~s$^{-1}$\,}
\newcommand{\masyr}{mas~yr$^{-1}$\,}
\newcommand{\msun}{$M_\odot$\,}

\title{Orbits of Six Triple Systems}

\author{Andrei Tokovinin}
\affiliation{Cerro Tololo Inter-American Observatory | NSF's NOIRLab
Casilla 603, La Serena, Chile}
\email{andrei.tokovinin@noirlab.edu}


\begin{abstract}
Joint analysis of  position measurements and radial  velocities of six
triple stellar  systems is conducted  to determine their  inner and/or
outer  orbits.  Accumulation of  such  data  is  needed to  study  the
architecture of stellar hierarchies and  its relation to the formation
mechanisms. The  inner periods in  the six systems (HIP  11783, 64836,
72423, 84720,  89234, and 105404)  range from 0.5  days to 44  yr. The
shortest  outer period  of  3.34 yr  is found  in  the compact  triple
HIP~105404  (BS~Ind).   The  resolved  triple  system  HIP  64836  has
comparable inner and outer periods (5  and 30 yr), placing it near the
limit of  dynamical stability,  while its quasi-circular  and coplanar
orbits suggest a  1:6 mean motion resonance. The periods  in HIP 89234
(44  and $\sim$450  yr)  are  also comparable,  but  the mutual  orbit
inclination is large, 54\degr. Masses  of the components are estimated
and each system is discussed individually.
\end{abstract}



\section{Introduction}
\label{sec:intro}

Many  stars in  the  solar neighborhood,  including  the nearest  one,
$\alpha$~Cen, are arranged  in hierarchical systems, believed  to be a
natural outcome of the star formation process. Almost every star could
belong  to  a stellar  system  in  its youth,  at  least  for a  while
\citep{Lee2019,Offner2023}. Young  dynamically unstable  systems decay
rapidly, but  stable hierarchies survive  for a long time,  bearing an
imprint of  the formation history  in their orbital  architectures.

Separations in stellar  systems span a huge range,  from contact pairs
to $\sim  10^4$ au and  more.  Their  discovery and study  is possible
through combination of diverse and complementary methods, as no single
technique  can  cover  the  full  range.   Space  missions  like  TESS
\citep{TESS}  and  Gaia \citep{Gaia1,Gaia3}  help  us  in finding  new
hierarchies, but suffer from various shortfalls when it comes to their
detailed study.  The  major hurdle is to reach  adequate time coverage
and sampling.  Dedicated ground-based  programs address this need.  In
this paper, I  use the results of high-resolution  imaging and radial
velocity (RV) monitoring to determine orbital parameters in six nearby
triple     systems;    it     continues    our     previous    effort
\citep{TL2017,TL2020,Trip2021,Trip2023}.

\begin{figure}
\epsscale{1.1}
\plotone{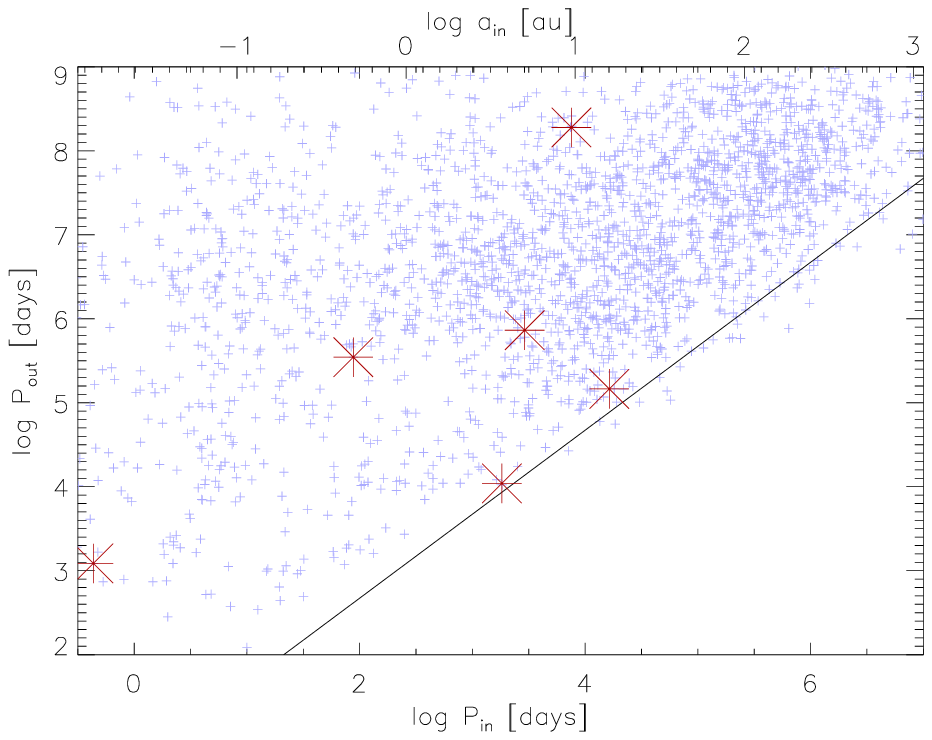}
\caption{Inner and outer periods  of known hierarchical systems within
  100\,pc (small crosses)  and the six triple systems  studied here (large
  asterisks). The line indicates the dynamical stability limit $P_{\rm
    out}/P_{\rm in}  > 4.7$  \citep{Mardling2001}.  Some  systems with
  crudely estimated periods are located below this line.
\label{fig:plps}
}
\end{figure}

\begin{deluxetable*}{c l cc cc c c l  }
\tabletypesize{\scriptsize}
\tablewidth{0pt}
\tablecaption{List of Multiple Systems
\label{tab:list} }
\tablehead{
\colhead{WDS} &
\colhead{Name} &
\colhead{HIP} & 
\colhead{HD} & 
\colhead{$V$} &
\colhead{$\varpi$\tablenotemark{a} } & 
\colhead{$\mu^*_\alpha$} &
\colhead{$\mu_\delta$} &
\colhead{Masses} 
 \\
\colhead{(J2000)} &        &  & &
\colhead{(mag)} & 
\colhead{(mas)} & 
 \colhead{(\masyr)} &
 \colhead{(\masyr)} &
 \colhead{(\msun)}
}
\startdata
02321$-$1515 &  TOK 382        & 11783  & 15798   & 4.75  & 37.46 H  & $-$72 & $-$125 & 1.45+0.69 \\
             & BD$-$15 447     & 11759  & 15767   & 8.74  & 37.19 G  &  $-$73 & $-$117 & 0.77 \\
13175+2024    & YSC 149         & 64836 & \ldots   & 11.06 & 17.45 G & $-$48  & 6    & (0.65+0.63)+0.59 \\
14485$-$1720  & BU 346A         & 72423  & 130412   & 7.37 & \ldots   & 43   & $-$54  & 1.21+0.48  \\
              & BU 346B         & \ldots & \ldots  & 7.90 & 20.05 G  & 36    & $-$46  & 1.10 \\
17191$-$4638  & 41 Ara          & 84720  & 156274  & 5.48 & 113.75 G & 1029 & 106     & 0.87 \\
              & LHS 445         & \ldots & \ldots  & 8.69 & 113.87 G & 953  & 140     & 0.60+0.42 \\
18126$-$7340  & HR 6751         & 89234  & 165259  & 5.85 & 23.68 G  & $-$91 & $-$254 & 1.47+0.73 \\
              &  HDO 284B       & \ldots & \ldots  & 9.28 & 23.71 G  & $-$62 & $-$241 & 0.82 \\ 
21210$-$5229  & BS Ind          & 105404 & 202947  & 8.91 & 19.00 D  & 40    & $-$94  & 0.85+(0.77+0.55) \\
\enddata
\tablenotetext{a}{Parallax codes: G --- Gaia DR3 \citep{Gaia3}, H --- Hipparcos \citep{HIP2}, D --- dynamical.}
\tablecomments{Explanation of columns: (1) WDS code \citep{WDS};
(2) WDS or other name
(3) Hipparcos number;
(4) HD number; 
(5) visual magnitude; 
(6) parallax;
(7) PM in R.A.; 
(8) PM in decl.; 
(9) masses.
 }
\end{deluxetable*}

Basic  parameters of  the triple  systems studied  here are  listed in
Table~\ref{tab:list}. These  stars are  relatively bright  and nearby,
within 100\,pc. To put  this study into context, Figure~\ref{fig:plps}
shows  inner and  outer periods  of known  hierarchies within  100\,pc
listed in the Multiple Star Catalog \citep[MSC,][]{MSC}.  Only a minor
fraction  of these  systems  have known  orbits  (unknown periods  are
estimated roughly from projected  separations).  This study contributes
ten inner and/or outer orbits in  six systems spanning a wide range of
periods (large asterisks).

The   input    data   and   methods   are    briefly   introduced   in
section~\ref{sec:methods}.  Sections \ref{sec:11783}--\ref{sec:105404}
are devoted to individual  systems, section~\ref{sec:sum} contains the
summary.

\section{Data and methods}
\label{sec:methods}

\subsection{Speckle Interferometry}
\label{sec:speckle}

In the  hierarchies studied here,  subsystems have been  discovered or
resolved  by speckle  interferometry with  the high-resolution  camera
(HRCam) working  on the  4.1 m   SOAR  (Southern Astrophysical
Research Telescope)  located in Chile.    The
instrument,   data  processing,   and  performance   are  covered   in
\citet{TMH10,HRCam}.  The latest series of measurements and references
to prior observations can be found in \citet{Tok2024}.  Image cubes of
200$\times$200 pixels  and 400 frames  are recorded mostly in  the $y$
(543/22\,nm) and  $I$ (824/170\,nm) filters  with an exposure  time of
25\,ms  and  a  pixel  scale  of 15\,mas.   In  the  $y$  filter,  the
difffraction-limited resolution  of 30\,mas can be  attained, and even
closer separations can be measured  via careful data modeling.  On the
other  hand, the  $I$ filter  offers a  deeper magnitude  limit and  a
better sensitivity to faint, red companions.

Image  cubes are  processed by  the standard  speckle method  based on
calculation of  the spatial power spectrum  and image auto-correlation
function (ACF)  derived from the  latter.  The 180\degr  ~ambiguity of
position angles inherent to this  method is resolved by examination of
the  shift-and-add (``lucky'')  images  and by  comparison with  prior
data. In a  triple star, the angles of subsystems  are related, so the
better-defined   orientation  of   the  outer   pair  constrains   the
orientation of the  inner subsystem.  

For the outer pairs discovered visually, position measurements at SOAR
are complemented by the historic micrometer and speckle data retrieved
from  the Washington  Double  Star (WDS)  database  \citep{WDS} on  my
request.

\subsection{Radial Velocities}
\label{sec:RV}

Radial velocities of multiple systems come from various sources.
Some stars were targeted by the CHIRON high-resolution optical echelle
spectrometer  at  the   1.5  m  telescope  located   at  Cerro  Tololo
\citep{CHIRON}.   The  spectra  with  a  resolving power  of  80,000  were
processed  by  the instrument  pipeline  and  cross-correlated with  a
binary   mask  based   on  the   solar  spectrum,   as  described   in
\citet{chiron1}.   The resulting  cross-correlation function  (CCF) is
approximated by  one or two Gaussians to  derive the
RVs and  other parameters, namely the  amplitude and width of  the CCF
dips.

Reduced spectra of  two targets obtained with the  FEROS fiber echelle
spectrograph \citep{FEROS} at the 2.2 m telescope were downloaded from
the ESO  science archive.  The  spectral resolution is  48,000.  These
spectra  were  cross-correlated  with  the  same  solar-type  mask  to
determine the RVs.

The RVs of HIP 72423 (HD 130412) have been monitored as part of the survey
of nearby solar-type stars conducted at the Harvard-Smithsonian Center
for  Astrophysics   (CfA)  using  several  instruments   on  different
telescopes. The  instruments and data reduction  methods are described
in our previous papers \citep{TL2017,TL2020}.

\subsection{Distances and Masses}
\label{sec:dist}

The  Gaia   data  release   3  \citep[DR3,][]{Gaia3}   gives  accurate
astrometry based  on 3 years  of data.   For close binary  stars, the
astrometry in DR3 is either missing or biased by the unmodeled orbital
motions.   Questionable  astrometry  is evidenced  by  large  parallax
errors  and large  values  of  the Reduced  Unit  Weight Error  (RUWE)
parameter.   The  bias is  reduced  for  the orbital  or  acceleration
solutions  in the  Gaia  non-single  star catalog  \citep{Arenou2022}.
Otherwise,  accurate parallaxes  of distant  non-binary components  of
triple systems should be used. In some cases, the shorter time base of
Gaia DR2 gives  less biased parallaxes because the  orbital motion was
closer to linear.

The  masses  of  the  components are  estimated  from  their  absolute
magnitudes $M_V$  using  standard relations  for main
sequence dwarfs \citep{Pecaut2013}.  The  combined flux is distributed
between the  unresolved components  in suitable proportion,  guided by
the magnitude differences measured  by speckle interferometry or by other
arguments.   This relative  photometry is  not very  accurate (typical
scatter of  0.1  mag  or  more  in the  $I$  band).   Lacking  accurate
differential photometry in  several bands, the adopted  flux ratios in
$V$ are only approximate; however, considering the strong dependence
of  $M_V$ on mass, the ``photometric'' masses are determined
reasonably well. They are checked against the mass sum derived from
well-determined orbits and parallaxes.

\subsection{Orbit Calculation}
\label{sec:orbit3}

As  in  the previous  papers,  an  IDL  code  {\tt orbit3}  that  fits
simultaneously inner and outer orbits  in a triple system to available
position      measurements     and      RVs     has      been     used
\citep{ORBIT3}.\footnote{Codebase:
  \url{http://dx.doi.org/10.5281/zenodo.321854}}    The   method    is
presented in  \citet{TL2017}.  The weights are  inversely proportional
to the squares  of adopted measurement errors which  range from 2\,mas
to   0\farcs05  and   more  \citep[see][for   further  discussion   of
  weighting]{Trip2021}.

Motion in  a triple system  can be  described by two  Keplerian orbits
only approximately, but  the effects of mutual dynamics  are too small
to be detectable with the current  data.  The code fits 14 elements of
both orbits  and the  additional parameter $f$  -- the  wobble factor,
ratio of  the astrometric wobble  axis to the  full axis of  the inner
orbit.  For  resolved triples, $f =  q/(1+q)$, where $q$ is  the inner
mass ratio.  When the inner subsystem is not resolved, measurements of
the outer  pair refer to  the photocenter of  the inner pair,  and the
wobble  amplitude corresponds  to a  smaller factor  $f^* =  q/(1+q) -
r/(1+r)$, where  $r$ is  the flux  ratio.  The  code {\tt  orbit4} can
accept a mixture of resolved and unresolved outer positions; it adopts
a  fixed  ratio  $f^*/f$,  specified  for  each  system  as  an  input
parameter.

To  avoid   the  degeneracy  of   outer  orbits  resulting   from  
insufficient  time coverage,  some elements  are fixed  to reasonable
values  that agree  with the  estimated masses.   The resulting  outer
orbits are only representative; however, they are  useful for the
assessment of mutual dynamics.

The elements of inner and outer  orbits in the selected triple systems
are given  in Table~\ref{tab:orb}  in standard  notation.  Considering
the uncertain nature of some outer  orbits, the formal errors of their
elements  are only  lower limits.   When  both positions  and RVs  are
available,  the  argument  of   periastron  $\omega_A$  is  chosen  to
represent the  RVs of the  primary component, and the  node $\Omega_A$
gives correct  positions of  the secondary;  adding 180\degr  ~to both
elements does not affect the positions.

Individual  positions,  their adopted  errors,  and  residuals to  the
orbits  are  listed  in  Table~\ref{tab:speckle},  available  in  full
electronically.  The systems are identified  by their WDS codes (their
HD   and  HIP   numbers   are  given   in  Table~\ref{tab:list})   and
components. Compared  to the published  HRCam data, the  positions are
corrected     for    the     small    systematics     determined    in
\citet{Tokovinin2022} and,  in a few cases,  re-processed.  The second
column indicates the subsystem; for  example, A,BC refers to the angle
and separation between  A and unresolved pair BC, while  A,B refers to
the position of  the resolved component B relative to  A.  The RVs and
their   residuals   to   orbits   are   listed   in   the   electronic
Table~\ref{tab:rv}.   Larger  errors  are  assigned  to  some  RVs  to
down-weight their  impact on the  orbit, as commented in  the following
sections.

\begin{deluxetable*}{ l  c rrr rrr r r r r  }
\tabletypesize{\scriptsize}
\tablewidth{0pt}
\tablecaption{Orbital Elements \label{tab:orb}}
\tablehead{
\colhead{WDS} &
\colhead{System} &
\colhead{$P$} & 
\colhead{$T  $} &
\colhead{$e$} & 
\colhead{$a$} & 
\colhead{$\Omega_A$} &
\colhead{$\omega_A$} &
\colhead{$i$}  &
\colhead{$K_1$} & 
\colhead{$K_2$} & 
\colhead{$V_0$} 
 \\
\colhead{HIP}   & &   
\colhead{(yr)} & 
\colhead{(yr)} &
\colhead{ } & 
\colhead{($''$)} & 
\colhead{(degr)} &
\colhead{(degr)} &
\colhead{(degr)} & 
\colhead{(\kms)} &
\colhead{(\kms)} &
\colhead{(\kms)} 
}
\startdata
02321$-$1515 &  Aa,Ab & 20.68  & 2015.621      &  0.854   & 0.3616      & 23.1    & 23.4 & 120.3 &  7.56 &  \ldots & $-$28.85  \\
11783        &        &  $\pm$0.670 &$\pm$0.055&$\pm$0.009 & $\pm$0.0049&$\pm$1.3 &$\pm$2.6&$\pm$1.3 & $\pm$0.25 & \ldots & $\pm$0.10  \\
13175+2024 & Aa,Ab    & 5.000  & 2022.84      & 0.285     & 0.050 & 292.6       & 6.6  & 134.0  & 6.99 & 7.33 & \ldots   \\ 
64836      &          & $\pm$0.052 &$\pm$0.08&$\pm$0.020 & $\pm$0.002&$\pm$4.1 &$\pm$9.1&fixed &  $\pm$0.34 & $\pm$0.35 & \ldots  \\
13175+2024 & A,B      & 30.00      & 2005.16 &  0        & 0.207     & 295.2   & 0     & 143.2   & 2.3 & \ldots     & $-$7.0    \\     
64836      &          & $\pm$0.28 &$\pm$1.08& fixed & $\pm$0.017&$\pm$2.6 & fixed &$\pm$1.6 & fixed  & \ldots & fixed  \\
14485$-$1720 & Aa,Ab &  7.948   & 2023.804  & 0.642  & 0.0270    & 262.4 & 164.6      & 40.0   & 4.17 & \ldots & \ldots \\
72423      &  & $\pm$0.017 &$\pm$0.034 & $\pm$0.036 & $\pm$0.0013 & $\pm$3.5 & $\pm$4.6 & fixed & $\pm$0.40 & \ldots &  \ldots  \\
14485$-$1720 & A,B &  2000   & 1368  & 0.262  & 4.60 & 125.0    & 337.1  & 75.1       & 1.40   & 2.15 & 1.55  \\
72423      &  & fixed &$\pm$40 & $\pm$0.026 & fixed  & $\pm$1.1 & $\pm$4.5 & $\pm$0.6 & fixed & fixed &  $\pm$0.06  \\
17191$-$4638  & Ba,Bb  & 0.24069 & 2015.903 & 0.773 & 0.0410  & 242.2 & 170.6 & 57.9 & \ldots  & \ldots & \ldots  \\  
\ldots        &        & fixed & $\pm$0.012 & fixed & $\pm$0.0024 & $\pm$14.8 & $\pm$31.1 & $\pm$4.9 & \ldots  & \ldots & \ldots  \\ 
17191$-$4638  & A,B  & 954.2  & 1908.2 & 0.816 & 12.752 & 320.2 & 150.3 & 35.2 & 1.956 & \ldots & \ldots   \\ 
84720  &             & $\pm$68.7 &$\pm$0.28& $\pm$0.007 & $\pm$0.548&$\pm$2.5 & $\pm$1.3 &$\pm$0.8 & $\pm$0.068 & \ldots & \ldots  \\
18126$-$7340 & Aa,Ab & 44.2   & 2027.22 &  0   & 0.373 & 105.2 & 0    & 83.5  & \ldots & \ldots & \ldots  \\
89234        &       &  $\pm$1.3 &$\pm$0.32& fixed & $\pm$0.008 &$\pm$0.3 & fixed &$\pm$0.4 & \ldots & \ldots & \ldots  \\
18126$-$7340 & A,B   & 450       & 2147.2    & 0.300 & 2.014  & 65.3   & 348.8 & 46.7    & \ldots & \ldots & \ldots  \\
89234        &       &  fixed & $\pm$21.8 & fixed & $\pm$0.051 &$\pm$9.1 & $\pm$23.0 &$\pm$1.6 & \ldots & \ldots & \ldots  \\
21210$-$5229 & A,B   & 3.3473 & 2020.583 & 0.627   & 0.0552  & 248.72  & 357.12 & 42.35  & 13.76 & \ldots  & 10.62  \\
105404       &       & $\pm$0.0015 &$\pm$0.007& $\pm$0.006 & $\pm$0.0005&$\pm$0.69 & $\pm$0.90 &$\pm$2.13 & $\pm$0.36 & \ldots & $\pm$0.06  \\
\enddata
\end{deluxetable*}

\begin{deluxetable*}{r l l rrr rr l}    
\tabletypesize{\scriptsize}     
\tablecaption{Position Measurements and Residuals (Fragment)
\label{tab:speckle}          }
\tablewidth{0pt}                                   
\tablehead{                                                                     
\colhead{WDS} & 
\colhead{Syst.} & 
\colhead{Date} & 
\colhead{$\theta$} & 
\colhead{$\rho$} & 
\colhead{$\sigma_\rho$} & 
\colhead{(O$-$C)$_\theta$ } & 
\colhead{(O$-$C)$_\rho$ } &
\colhead{Method\tablenotemark{a}} \\
 & & 
\colhead{(JY)} &
\colhead{(\degr)} &
\colhead{(\arcsec)} &
\colhead{(\arcsec)} &
\colhead{(\degr)} &
\colhead{(\arcsec)} &
}
\startdata
 02321$-$1515 &Aa,Ab & 2013.7461 & 162.0 &  0.2090 &   0.005 & $-$0.2 &   0.000 & S \\
 02321$-$1515 &Aa,Ab & 2014.0410 & 157.4 &  0.1774 &   0.005 & $-$0.1 &  0.000 & S \\
 02321$-$1515 &Aa,Ab & 2014.0410 & 158.2 &  0.1768 &   0.005 &   0.7 &  $-$0.001 & S \\
 02321$-$1515 &Aa,Ab & 2016.9572 & 214.9 &  0.2389 &   0.005 &   0.8 &   0.001 & S 
\enddata 
\tablenotetext{a}{Methods:
G: Gaia;
H: Hipparcos;
M: visual micrometer measurement;
P: photographic measurement;
S: speckle interferometry at SOAR;
s: speckle interferometry at other telescopes.
}
\end{deluxetable*}

\begin{deluxetable*}{r l c rrr l }    
\tabletypesize{\scriptsize}     
\tablecaption{Radial Velocities and Residuals (Fragment)
\label{tab:rv}          }
\tablewidth{0pt}                                   
\tablehead{                                                                     
\colhead{WDS} & 
\colhead{Comp.} & 
\colhead{JD} & 
\colhead{RV} & 
\colhead{$\sigma$} & 
\colhead{(O$-$C)$$ } &
\colhead{Instr.\tablenotemark{a}}
 \\
 & & 
\colhead{ (JD $-$24,000,000)} &
\multicolumn{3}{c}{(km s$^{-1}$)}  &
}
\startdata
02321$-$1515 &Aa  &  52594.500  &  $-$29.50 &    2.00 &    0.64 & T2005 \\
02321$-$1515 &Aa  &  52597.500  &  $-$29.00 &    2.00 &    1.14 & T2005  \\
02321$-$1515 &Aa  &  52600.500 &  $-$29.80 &    2.00 &    0.34 & T2005  \\
02321$-$1515 &Aa  &  54706.6040 &  $-$29.45 &    0.50 &   $-$0.17 & RAVE  \\
02321$-$1515 &Aa  &  56908.8810 &  $-$24.58 &    0.10 &    0.00 & CHI \\ 
02321$-$1515 &Aa  &  57389.0000 &  $-$26.82 &    3.89 &   $-$0.33 & DR2 
\enddata 
\tablenotetext{a}{Instruments:
CHI: CHIRON;
CfA: CfA digital speedometers;
COR: CORAVEL;
DR2: Gaia DR2;
G2005: \citet{Guenther2005};
FEROS: FEROS;
RAVE: \citet{RAVE};
TRES: TRES;
T2005: \citet{Takeda2005}.
}
\end{deluxetable*}

\section{HIP 11783 ($\sigma$~Cet)}
\label{sec:11783}

\begin{figure}
\epsscale{1.1}
\plotone{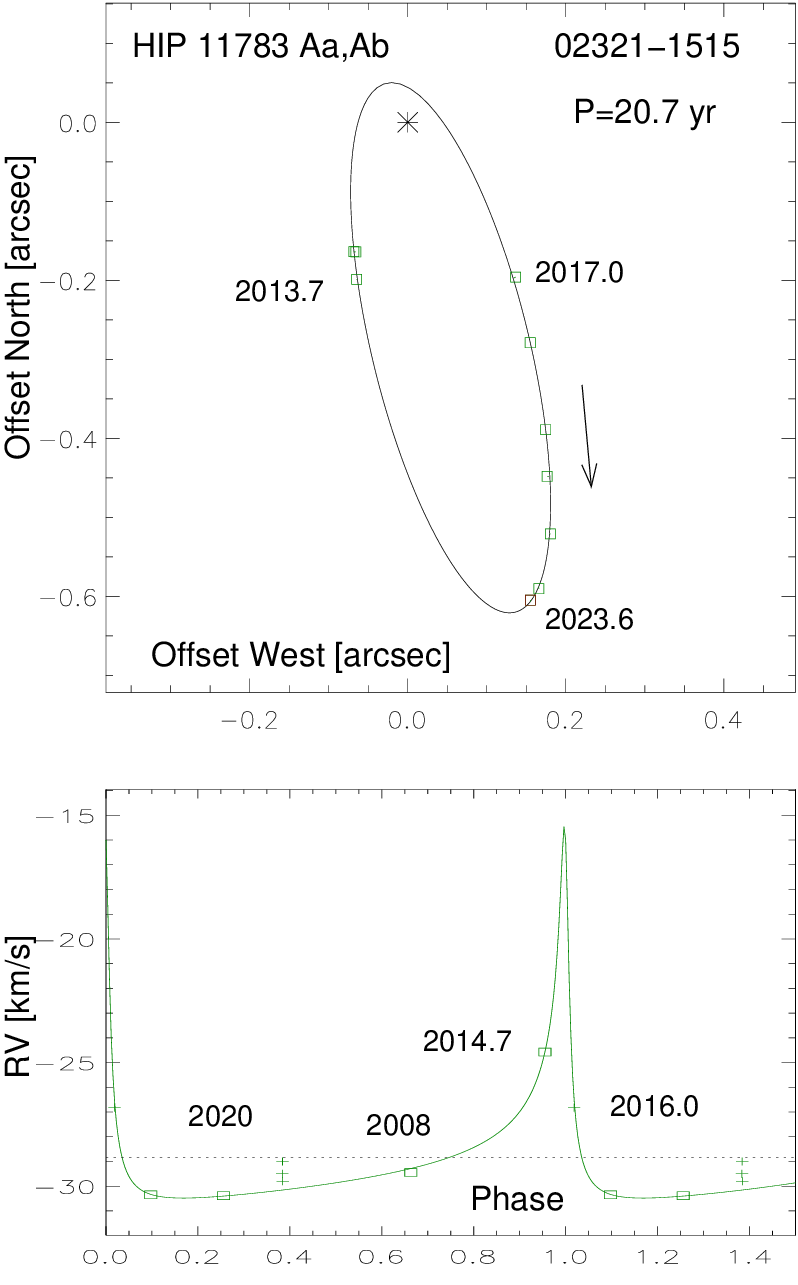}
\caption{Orbit of  HIP 11783 Aa,Ab in  the plane of the  sky (top) and
  its RV  curve (bottom).  Squares  plot the measurements,  the curves
  depict the  orbits. Crosses in the  RV curve mark the  less accurate
  RVs.
\label{fig:11783}
}
\end{figure}

This triple system is located  at 27\,pc distance.  The bright primary
component A ($\sigma$~Cet,  HR 740, HIP 11783, $V=4.75$  mag, F5V) and
star B (HIP~11759, HD 15767,  $V=8.74$ mag, K2.5V) at 345\arcsec ~form
a wide common proper motion (CPM)  pair designated as GWP~335AB in the
WDS  with an  estimated period  of $\sim$500  kyr.  The  most accurate
distance to the  system results from the Gaia DR3  parallax of star B,
37.19$\pm$0.02 mas.  It matches the HIP2 \citep{HIP2} parallax of star
A,  37.46$\pm$0.25 mas,  while the  Gaia astrometry  of A  is strongly
biased by its orbital motion near  the periastron (e.g.  a parallax of
43.38$\pm$0.56 mas in DR3). The PM of A listed in Table~\ref{tab:list}
is estimated by \citet{Brandt2018} from positions in Hipparcos and
Gaia DR2; it is  slightly biased, but closer to reality than the Gaia
PM. 

The  bright star  A  has an  extensive  literature. The  spectroscopy,
summarized in  \citet{PASTEL}, indicates  an effective  temperature of
6300\,K, a surface gravity $\log g$  from 3.7 to 4.0 (slightly evolved),
and a metallicity of $-0.30$ to $-0.25$ dex. Star A is known to have a
variable  RV \citep{N04}  and  astrometric acceleration  \citep{MK05}.
Its faint  companion Ab  was resolved  for the first  time at  SOAR in
2013.7 at 0\farcs2 separation with a large contrast (average $\Delta I
= 3.89$  mag with  an rms scatter  of 0.17 mag);  it is  designated as
TOK~382Aa,Ab in the WDS.

The RVs of A  and B were measured in 2014.7 using  CHIRON and found to
be  substantially  different.   At  that  time,  the  pair  Aa,Ab  was
approaching periastron of its eccentric orbit and it became unresolved
at  SOAR  after  2014.   The  RVs and  positions  measured  after  the
periastron allowed  calculation of  an orbit  with a  period of  21 yr
\citep{Tok2021c}.  The orbital elements  are updated here using recent
speckle  measurements   and RVs (Figure~\ref{fig:11783}).   Despite the
large contrast, the  rms residuals to the orbit are  less than 2\,mas.
Crosses on  the RV  curve denote  RVs used with  a low  weight, namely
three points from \citet{Takeda2005} and the mean RV in Gaia DR3.  The
latter has a large error of 3.89  \kms owing to the rapid RV variation
near periastron.  One  RV from RAVE DR6 \citep{RAVE}  measured in 2008
is also used.  When the RV  series from Gaia become publicly available
in DR4, they can be used to  improve the orbit; the next periastron in
2036 will present another chance to cover it spectroscopically.

The Gaia DR3  parallax of B and  the orbit of Aa,Ab  correspond to the
mass sum of 2.15 \msun. Adopting the mass of 1.45 \msun for Aa, the RV
amplitude  and inclination  lead  to  a mass  of  0.70  \msun for  Ab,
matching the mass sum. The mass  of the CPM companion B estimated from
its absolute magnitude is 0.77 \msun.  Its mean RV of $-28.23$ \kms is
very close  to the center-of-mass  velocity of A, $-28.85$  \kms.  The
CHIRON spectrum  of B  shows no  lithium 6707\AA  ~line, while  in the
spectrum of A it is  prominent (equivalent width 78\,m\AA).  The axial
rotation of A estimted  from the CCF width is moderate,  $V \sin i =7$
\kms.



\section{HIP 64836 (YSC 149) }
\label{sec:YSC149}

\begin{figure}
\epsscale{1.1}
\plotone{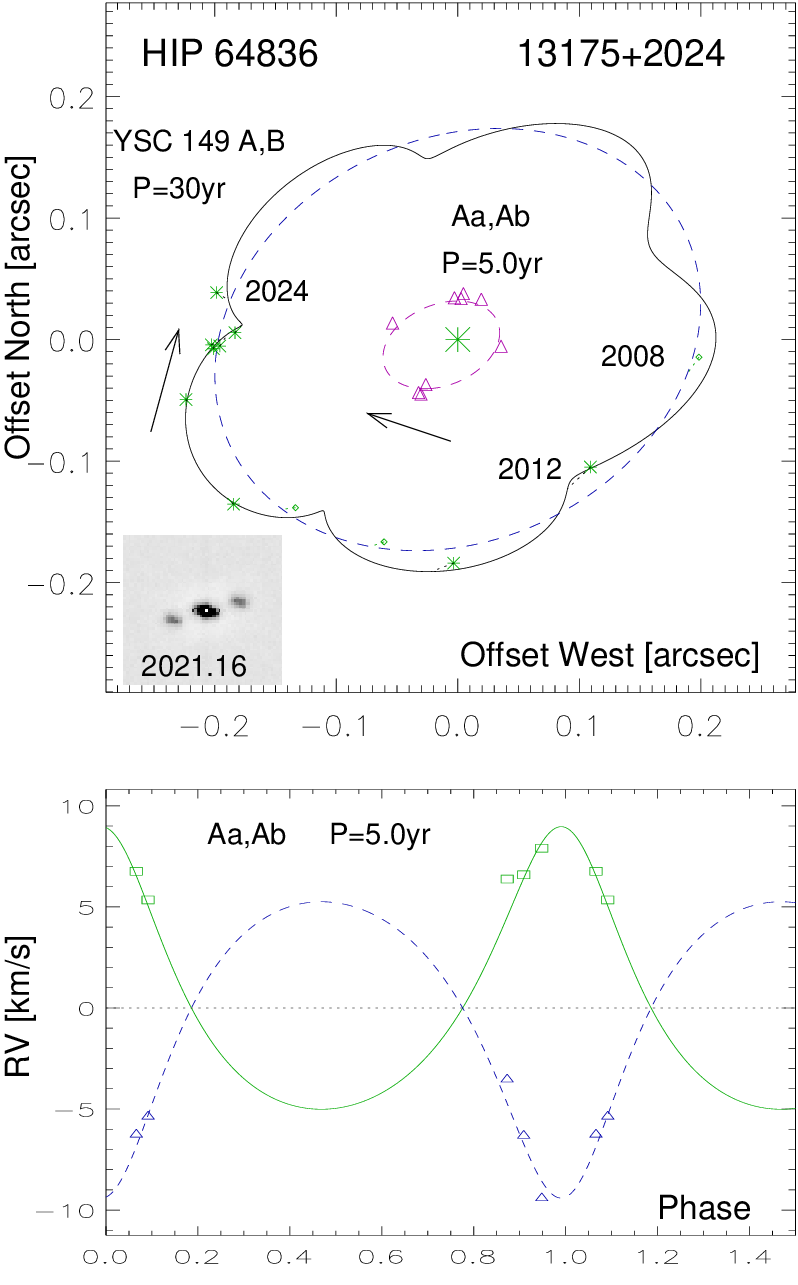}
\caption{Orbits of HIP  64836 (YSC 149) in the plane  of the sky (top)
  and  the RV  curve of  the inner  pair (bottom,  the outer  orbit is
  subtracted). In  the orbit  plot, asterisks, crosses,  and triangles
  show  the resolved  outer,  unresolved outer,  and inner  positions,
  respectively.  The  inner  orbit  is plotted  by  the  small  dashed
  ellipse. The outer orbit is plotted by the solid line (with wobble)
  and  by the  dashed line  (with wobble  removed).  The  insert shows  the
  speckle ACF recorded in 2021.
\label{fig:64836}
}
\end{figure}

The  K5V dwarf  HIP 64836  (WDS  J13175+2024, $V=11.06$  mag)  is
located  in the  solar  vicinity (GJ  9436, MCC  687).   The Gaia  DR3
parallax  is 17.45$\pm$0.45  mas, with  a RUWE  of 30.   The Hipparcos
\citep{HIP2}  gives   a  stochastic   solution  with  a   parallax  of
16.69$\pm$2.57\,mas.  \citet{Horch2012} resolved  this star in 2008.47
into  a  0\farcs2  pair  YSC~149.  Their  subsequent  observations  in
2012.09  and  2014.46 revealed  this  as  a  triple system  where  the
secondary component is a close pair  Ba,Bb with a separation of 50 mas
\citep{Horch2015,Horch2017}.

Observations of this  system at SOAR in 2016, 2018,  and 2019 produced
measurements of  the A,B  pair without  resolving the  subsystem.  The
latter  was clearly  resolved in  2021.16 and  marginally resolved  in
2022--2024. The SOAR  data indicate that the subsystem  belongs to the
primary: it is  Aa,Ab rather than Ba,Bb. This  conclusion is supported
by the relative  photometry. Star B is  fainter than A by  1.43 mag in
the $V$ band  \citep{Horch2012} and by 1.1 mag in  $I$ (SOAR). Yet, in
2012  Horch  et al.   measured  magnitude  differences  of Ba  and  Bb
relative to  A of 0.62  and 0.67 mag  (at 692\,nm), implying  that the
combined light of  Ba and Bb exceeds the light  of A. Their photometry
in 2014 again suggests that B is brighter  than A. If B were a pair of
similar stars, each of them should be $\sim$1.8 mag fainter than A.

Attributing the subsystem to A leads  to the $V$ magnitudes of Aa, Ab,
and B of 11.97, 12.18, 12.42 mag (assuming $\Delta V_{\rm A,B} = 1.43$
mag and $\Delta V_{\rm Aa,Ab} = 0.2$ mag).  Given the DR3 parallax and
the  standard   mass-luminosity  relation   \citep{Pecaut2013},  these
magnitudes  correspond to  the masses  of 0.65,  0.63, and  0.59 \msun
(mass sum 1.85  \msun).  All three stars in this  system are therefore
 late-K dwarfs.

\begin{figure}[ht]
\epsscale{1.1}
\plotone{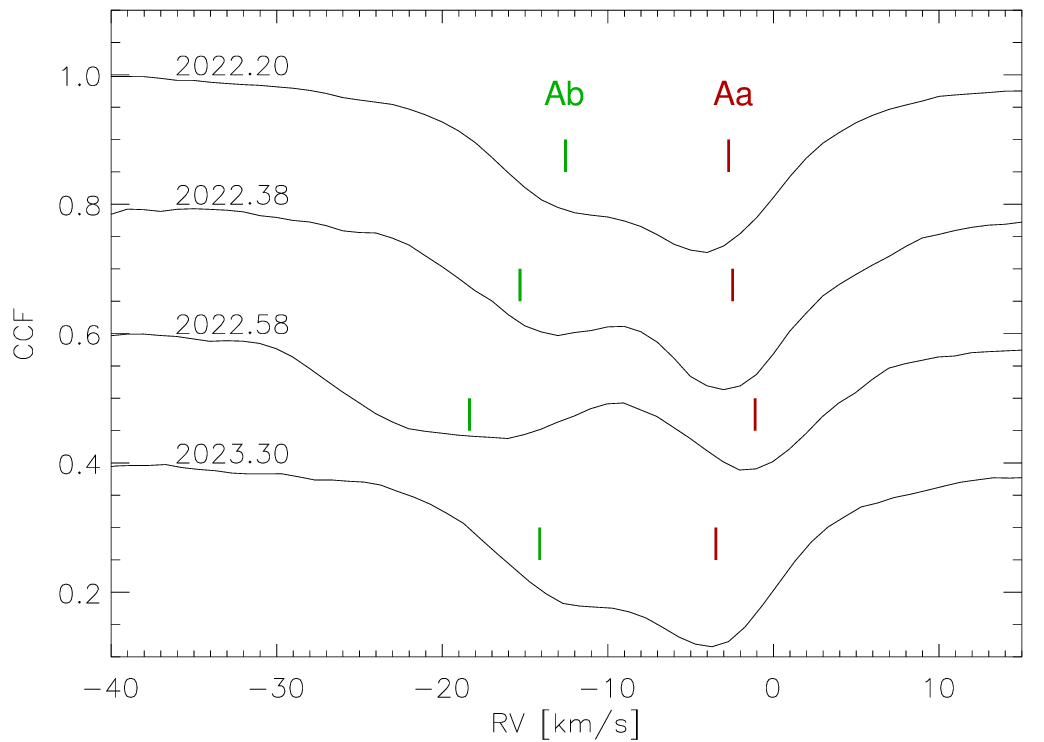} 
\caption{Four CCFs of  HIP 64836 derived from the  CHIRON spectra. The
  dates  are   indicated.  The  color  ticks  mark  the  measured
  heliocentric RVs of Aa (red, on the right) and Ab (green).
\label{fig:ccf} 
}
\end{figure}

The separations  imply the inner and  outer periods of the  order of 5
and 30 yr,  respectively. About half of the outer  orbit is covered. I
averaged the positions  of A,Ba and A,Bb reported by  Horch et al.  to
get the ``unresolved'' positions of A,B and fitted a preliminary outer
orbit. Independently,  B.~Mason computed a preliminary  outer orbit of
YSC~149 with a period of 32.96 yr \citep{Tok2024}.
Guessing  the orbit  of the  inner  subsystem Aa,Ab  was less  obvious
because  the  period  is  short, the  magnitude  difference  is  small
($\Delta I_{\rm Aa,Ab} = 0.15$ mag  with rms scatter of 0.11 mag), and
the separation  of Aa,Ab  is always near  the diffraction  limit. This
allows swaps between Aa and Ab.   The SOAR observations in 2016.13 and
2019.38  were  reprocessed  using  the triple-star  model.   The  data
suggest  retrograde motion of the inner pair with a period of $\sim$5
yr.

The  estimated RV  amplitude in  the inner  pair is  $\sim$5 \kms,  so
triple lines  could possibly  be detected in  high-resolution spectra.
The   resolution   of   three   CORAVEL   observations   reported   by
\citet{Sperauskas2016}  is not  high  enough to  detect triple  lines.
Four  spectra were  taken  with CHIRON.   The  CCFs look  double-lined
(Figure~\ref{fig:ccf}). The stronger  dip on the right  side (red tick
marks) is identified here with Aa, and the left-side dip (green ticks)
with Ab.  Against  expectation, the CCF is  not triple-lined, possibly
because the  weakest star B is  a fast rotator with  shallow lines.  I
recovered from the  ESO science archive one spectrum  taken with FEROS
in 2023 March  (JD 2,460,007.7544, program led  by E.~Costa).  Another
FEROS  spectrum taken  by E.~Costa  on my  request in  2024 April  (JD
2,460,420.651) shows  only one blended CCF  dip with an RV  of $-7.33$
\kms which is not used in the orbit fit, but matches its prediction.

The inner and  outer positions and RVs were fitted  jointly using {\tt
  orbit4}. Regarding the alternative outer positions measured by Horch
et  al.  in  2012 and  2014, I  selected those  that match  the joint
orbits   (Table~\ref{tab:orb}).   Figure~\ref{fig:64836}   shows  both
orbits. The  resolved outer positions  Aa,B are plotted  as asterisks,
the unresolved A,B  as small crosses, and the dashed  line shows the
outer obit without wobble.  The fitted wobble factor $f=0.41 \pm 0.05$
corresponds to the inner mass ratio $q_{\rm in}=$0.75$\pm$0.15, the RV
amplitudes   indicated   $q_{\rm   in}=   0.71$,   contradicting   the
near-equality of  stars Aa and Ab.   A closer examination of  the data
shows that the  individual RV amplitudes of Aa and  Ab are constrained
weaker than  their sum, so both  the spectroscopic mass ratio  and the
wobble amplitude are  somewhat uncertain.  In the  final iteration, I
fixed  $f=0.49$, the  outer RV  amplitude $K_3  = 2.3$  \kms, and  the
center  of mass  velocity $V_0  = -7.0$  \kms to  get nearly  equal RV
amplitudes  in  the  inner  orbit.   The free  fit  yields  the  inner
inclination of  $i_{\rm in} = 135\fdg1  \pm 5\fdg3$, so $i_{\rm  in} =
134\degr$  ~was fixed  to match  the RV  amplitudes with  the expected
inner  mass  sum  of  1.28  \msun.   The  outer  eccentricity  is  not
significantly different from zero, so  $e_{\rm A,B} = \omega_{\rm A,B}
= 0$ were also fixed.

Overall, these orbits give a  good match to the available observations
and to the properties  of normal low-mass stars.   The mutual inclination
between the orbits is $\Phi = 10\fdg0 \pm 1\fdg7$; the period ratio is
$6.0 \pm  0.1$, hinting at a  6:1 mean motion resonance.   This triple
system  resembles   some  other  low-mass  hierarchies   with  similar
architecture (quasi-circular  and coplanar orbits,  comparable masses,
small  period ratio),  such  as the   triple system
00247$-$2653  or LHS~1070
\citep{Dancingtwins}, where the latest  speckle data correspond to the
inner  and  outer  periods  of  17.26$\pm$0.02  and  83.0$\pm$1.8~yr,
respectively, and their ratio is 4.8$\pm$0.1.

For  a test,  the outer  orbit was  fitted separately  by using  outer
positions corrected  for the inner subsystem,  the mean RVs of  Aa and
Ab, and the  less accurate blended RVs  from \citet{Sperauskas2016} to
extend  the time  coverage.  This  experiment confirmed  the near-zero
outer eccentricity and the adopted ascending node of the outer orbit.

The Gaia DR3  parallax of 17.45$\pm$0.51 mas corresponds  to the inner
and outer  mass sums of 0.95  and 1.85 \msun, respectively.   The Gaia
astrometry  is  affected  by  the complex  motion  of  the  unresolved
photocenter.  Fitting  Gaia transits  by an  acceleration or by a single
Keplerian orbit will not be  sufficient.

\section{HIP 72423 (HD 130412, BU 346)}
\label{sec:BU 346}

\begin{figure}
\epsscale{1.1}
\plotone{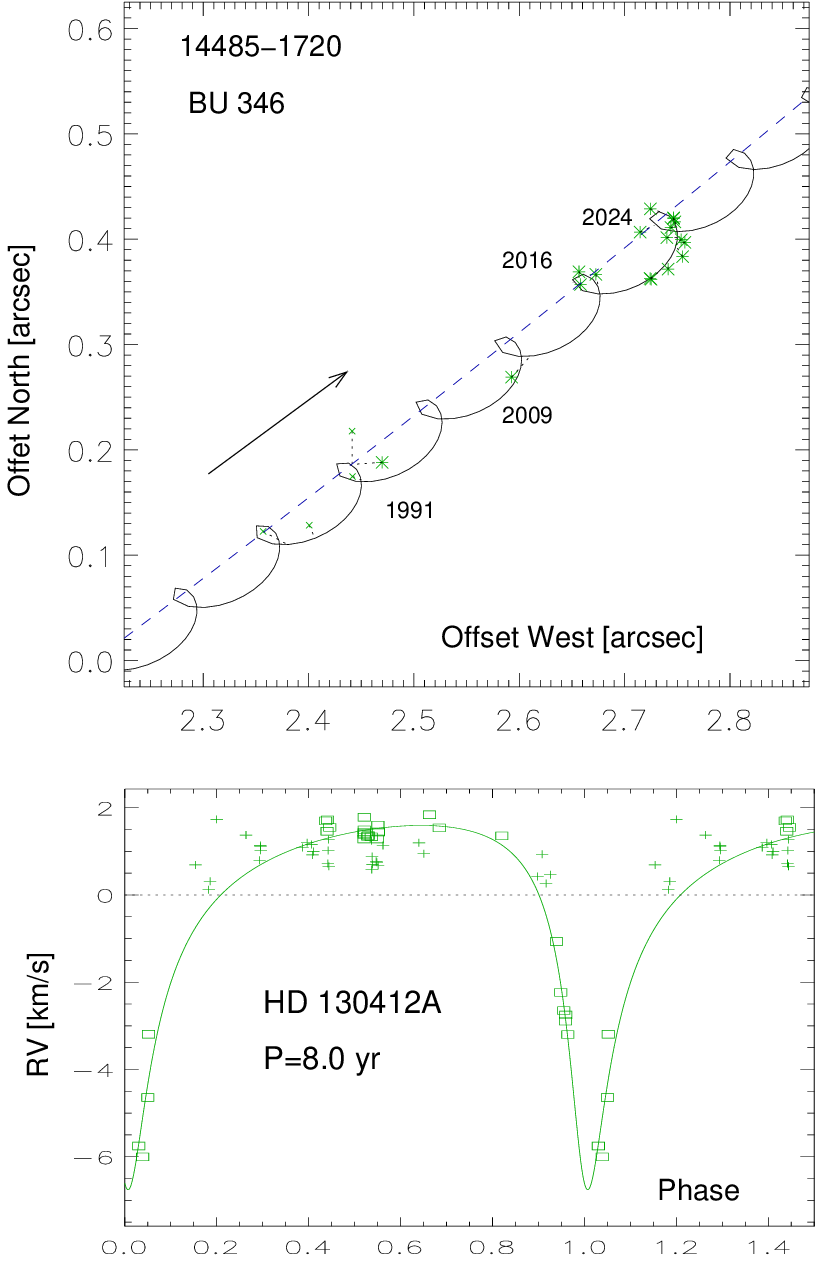}
\caption{Top:  fragment of  the  outer orbit  with wobble.   Accurate
  measurements (HIP, Gaia, SOAR) are plotted as asterisks, less accurate
  data  by  crosses.   The  dashed  line  depicts  the  center-of-mass
  trajectory.  Bottom: the  RV curve of star A.  Plus  signs denote RVs
  potentially biased by blending; the center-of-mass RV is subtracted.
\label{fig:orbBU346}
}
\end{figure}


The visual pair BU~346 (ADS 9387, WDS J14485$-$1720) has been resolved
in 1875 at  a separation of 1\arcsec \citep{Burnham1875}.   A slow and
almost linear  orbital motion was  revealed during 1.5  centuries, and
now the  separation has increased  to 2\farcs75.   Stars A and  B have
individual entries  in the Gaia  DR3 catalog, which gives  no parallax
for A and  a parallax of 20.046$\pm$0.025 mas for  B.  The DR2 catalog
contains astrometry for both A and B, but the nonlinear motion of star
A  definitely  disagrees  with  the Gaia  default  single-star  model,
explaining the absence of its parallax  and proper motion (PM) in DR3.
Gaia gives $G$ magnitudes of 7.23 and 7.72  mag for A and B (or $V$ of
7.37 and 7.86 mag).   The masses of Aa and B are  estimated as 1.2 and
1.1  \msun  from their  $V$-band  absolute  magnitudes using  the  DR3
parallax of B.


The RV of A has been monitored  by the CfA team led by D.~Latham since
1984 and  found to  be variable.   According to  the private  notes, a
tentative orbit with $P=4.7$ yr has been determined but not published.
After  preliminary analysis  of all  available  data in  2023 May,  it
became  clear   that  the   inner  subsystem  Aa,Ab   was  approaching
periastron, and a  fast RV variation was expected.  In 2023 May--June,
the system was observed twice with CHIRON and four times with the TRES
spectrograph  \citep{Szentgyorgyi:2007}.   A  negative  RV  trend  was
revealed, confirming the predicted periastron in 2023.7.

An  orbit  of A,B  with  $P=1715.7$  yr  and $a=9\farcs407$  has  been
determined by \citet{Izmailov2019} via formal least-squares fit to all
measures  collected in  the  WDS. With  a parallax  of  20.05 mas,  it
corresponds to the  mass sum of 35  \msun, hence it is  unlikely to be
correct. The outer orbit is  poorly constrained by the short (42\degr)
observed  arc.  The  pair  A,B was  originally included  in  a set  of
calibrator binaries for SOAR speckle interferometry.  For this reason,
it was  observed rather  frequently.  Accurate SOAR  measurements show
clearly a deviation  from the linear motion (wobble).   I assume that
the  wobble  is  produced  by   the  subsystem  in  the  component  A,
considering its variable RV.

The IDL code {\tt  orbit3} was used to simultaneously fit the positions
of   A,B    and   the   RVs    of   A   by   two    Keplerian   orbits
(Figure~\ref{fig:orbBU346}).  The outer period and semimajor axis were
fixed to  values that give the  estimated mass sum of  2.8 \msun.  The
inner period was found to be  about 8 yr.  Inaccurate speckle measures
by Horch on  a 0.8 m telescope and the Washington  speckle data from the
26  inch  refractor  are  ignored,  as well  as  many  micrometer  and
photographic  measures.  The  inner  pair is  never  resolved, so  the
wobble factor is fixed to $f=1$, and the resulting $a_{\rm in}$ is the
astrometric axis.

D.~Latham has communicated  the RVs of A and B  measured with CORAVEL,
CfA  digital spectrometers,  and TRES,  corrected  to be  on the  same
system (add +0.14 \kms to bring these RVs to the absolute system).  In
the  final iteration  of the  orbits, these  RVs were  used.  However,
considering the  small separation between  A and  B on the  sky, their
light can be frequently blended, biasing the RVs of A toward the RV of
companion B  (its mean RV  is +1.7  \kms).  Attempting to  resolve the
visual  pair, observers  likely offset  the object  on the  slit, also
creating an  RV bias.   To avoid subjective  selection, I  assigned an
increased error of 0.9  \kms to all CfA RVs between  0.6 and 2.6 \kms,
assuming that they  can be biased by blending; the  remaining RVs were
given the  unmodified errors  around 0.35  \kms. Some  RVs of  star B,
apparently affected by blending or  misidentification with A, are also
strongly  down-weighted. The  RVs measured  with TRES  and CHIRON  are
assigned errors of 0.2 \kms.  The  weighted residuals to the orbit are
0.35 \kms for Aa and 0.26 \kms for B.  The weighted position residuals
are 4 and 5\,mas  in X and Y and also match  the assigned errors.  The
SOAR measurement errors are  dominated by the calibration uncertainty.
Figure.~1 in  \citet{Tokovinin2022} suggests that the  position errors
of the calibrator binaries of similar separation are about 4\,mas.

The free  fit leads to a  small inclination of the  inner orbit, which
would imply an unrealistically large mass of the companion Ab. So, the
inner  inclination was  fixed at  40$^\circ$, yielding  a mass  of 0.5
\msun for Ab. The joint fit  of positions and RVs constrains the inner
orbit much better than each of these data sets alone.  The astrometric
axis  is 27.0\,mas,  while the full  axis of  the Aa,Ab
orbit, computed from the mass sum  and period, is 95.1\,mas. Their ratio
gives the  wobble factor $f=0.28$  and the mass ratio  $q_{\rm Aa,Ab}=
0.40$;  if the mass  of Aa is  1.21 \msun, the  mass of Ab  is 0.48
\msun. 

Given that the data do not  constrain well the outer orbit, a circular
orbit  with $P=1900$\,yr  was fitted  initially to  the observed  arc.
However, it predicted a substantial  (about 3 \kms) difference between
the systemic  RV of A and  the RV of B,  while they agree to  within 1
\kms.  The outer  inclination is certainly large,  while the estimated
RV amplitudes in  the outer orbit are  1 to 2 \kms.  So,  the small RV
difference between A  and B indicates that the outer  pair is close to
its  node. The  final  solution  with an  outer  eccentricity of  0.26
matches  this additional  constraint.   The outer  RV amplitudes  were
fixed to  their values estimated from  the adopted masses of  A and B,
1.7 and 1.1 \msun, and the  joint fit was repeated several times.  The
results reported  here refer to  the final iteration. The  outer orbit
remains quite uncertain,  but its elements describe  well the observed
motion.

\citet{Brandt2018} detected a very large PM anomaly (PMA) of star A by
comparing its Gaia DR2  astrometry with Hipparcos. However, considering
that Gaia  was affected by the  inner orbit of star  A, its comparison
with the orbits  can be only qualitative.  If we  take the accurate PM
of star  B measured  in DR3, $(36.2,-47.0)$  \masyr, and  subtract the
motion in the  outer orbit, $(-9.9,+7.5)$ \masyr, the  resultant PM of
star A,  $(46.1, -54.5)$ \masyr,  is in reasonable agreement  with the
average DR2$-$HIP  PM of  $(43.2,-54.2)$ \masyr determined  by Brandt.
Deviations of the  PM of A from  its average PM in  1991.25 and 2015.5
predicted by the inner orbit approximately match the measured PMs.

The mutual  inclination between orbits  $\Phi$, defined by  the angles
$\Omega$  and  $i$, is  known  better  than  the uncertain  period  or
eccentricity of the outer orbit. It  is $\Phi = 107 \pm2^\circ$ if the
ascending node of the outer orbit is chosen correctly on the basis
of the  small RV  difference between  A and  B.  The  alternative node
corresponds to a mutual inclination of  $48^\circ$. Therefore, to the best of
our  current knowledge,  the inner  and  outer orbits  in this  triple
system are almost orthogonal.

\vspace{0.5cm}

\section{HIP 84720 (41 Ara)}
\label{sec:84720}


\begin{figure}
\epsscale{1.1}
\plotone{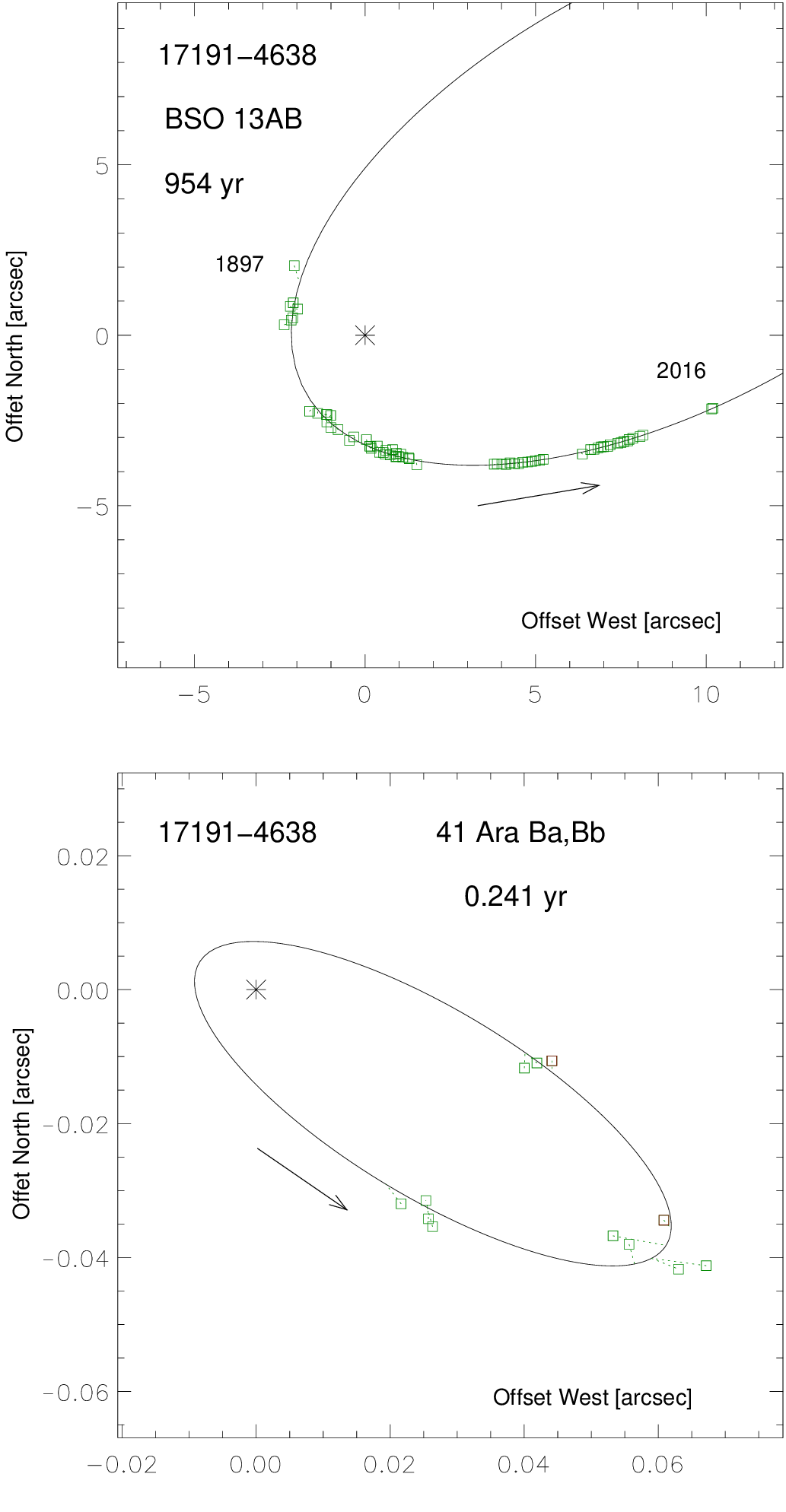}
\caption{Orbits of the outer (top) and inner (bottom) pairs in 41 Ara.
\label{fig:41Ara}
}
\end{figure}

The bright star 41 Ara (HR 6416, GJ~666A, WDS J17191$-$4638, G8/K0V) is
located  at  8\,pc  from  the  Sun. Just  like  our  nearest  neighbor
$\alpha$~Cen, this  is a triple system.  The outer pair A,B  was first
resolved at the end of the  19th century at 3\arcsec ~and presently it
has opened up  to 10\arcsec, moving slowly on  a millenium-long orbit.
The secondary star  B was announced as an 88  day spectroscopic binary
by   \citet{R10},  but   its  spectroscopic   orbit  has   never  been
published. Recently, Gaia determined an astrometric orbit of this pair
with  the same  period.   The  Gaia orbital  solution  also gives  the
unbiased parallax of 113.86$\pm$0.03  mas, in excellent agreement with
the parallax of  star A (113.75 mas), and the  unbiased PM.  The Ba,Bb
pair was tentatively resolved at SOAR  in 2016 and was monitored since
that time.  Despite being relatively  bright, it is a difficult target
for speckle interferometry because the magnitude difference in the $I$
band is substantial,  1.41$\pm$0.11 mag, and the  separation is always
close to the  diffraction limit. The orbit of Ba,Bb  based on the SOAR
data was announced in \citet{Circ209}; it is updated here.  The period
and  eccentricity  are  fixed  to   the  values  determined  by  Gaia.
Figure~\ref{fig:41Ara} plots the inner and outer orbits.


Although the  arc of the  outer orbit  covered by the  measurements is
relatively  large  and includes  periastron,  the  orbit is  not  well
constrained because several its elements  are strongly correlated. Several
orbits of A,B are found in the literature. Here only the more accurate
photographic measures, available since 1931, are used, complemented by
the less accurate earlier micrometer measures and the precise relative
positions from Hipparcos and Gaia.  This  orbit is very similar to the
953 yr orbit computed by  \citet{Scardia2013d}. On the other hand, the
latest 609 yr orbit by \citet{Izmailov2019}, obtained via a formal fit
to all data, is questionable.
The estimated  mass sum of A,B  is 1.88 \msun (see  below), larger but
compatible with the  mass sum deduced from the  parallax of 113.8\,mas
and the outer orbit (1.54 \msun).  If the outer eccentricity is forced
to a slightly smaller value, e.g.  0.75, the mass sum of A,B increases
and  approaches the  photometric  estimate, while  the  period of  A,B
becomes shorter.


\citet{Jenkins2015} detected  an RV trend  in the star A  from precise
measurements with the HARPS and UCLES spectrometers. I used the latter
data (provided  by D.~Rabout) because  they cover a larger  time span.
Their RVs (in m~s$^{-1}$) are determined with an arbitrary zero point.
The  fitted  RV amplitude  of  A,  1.956$\pm$0.067 \kms,  matches  the
expectation and  defines the true  ascending node of the  outer orbit,
while  the  systemic  velocity   is  meaningless.   The  RV  amplitude
corresponds to  a mass of  0.97 \msun for B,  assuming that A  is 0.87
\msun  (see  below).   So,  the  RV   trend  of  star  A  detected  by
\citet{Jenkins2015} is caused  by its motion in the  outer orbit.  The
rms residuals  to the orbit  are 4.3  m~s$^{-1}$, while the  quoted RV
errors are around 1 m~s$^{-1}$.   The additional RV scatter, if proven
to be real, can be caused by planets revolving around star A.  Lacking
permission  from the  authors, I  do  not reproduce  the RVs  of A  in
Table~\ref{tab:rv}.

Adopting the magnitude  difference $\Delta V_{\rm Ba,Bb}  = 1.75$ mag,
I get  the $V$  magnitudes of  Ba and Bb  of 8.89  and 10.64  mag and
estimate  the  masses  as  0.60  and  0.41  \msun  from  the  standard
main-sequence relations (the spectral type of B is M0VpCa-3Cr-1).  The
inner orbit corresponds to the mass  sum of 0.81 \msun.  The estimated
masses of Ba and Bb imply the mass ratio $q_{\rm Ba,Bb}=0.68$.  With a
magnitude difference of $\sim$1.5 mag in the $G$ band, the light ratio
is $r_{\rm Ba,Bb}=0.25$.  The  ratio of the astrometric amplitude,
9.3\,mas  (Gaia), to  the full  axis  of Ba,Bb,  41.0\,mas, gives  the
wobble factor $f^*  = 0.23$, in excellent agreement  with its estimate
$f^* = q/(1+q) - r/(1+r) = 0.21$.

The inner  semimajor axis is measured  with a relative error  of 0.06,
which    translates   to    the   relative    mass   sum    error   of
$\sim$0.2. Therefore, the constraint on the mass sum of Ba,Bb is weak,
and this binary  definitely cannot serve as a  benchmark. However, the
star  is   bright  in  the   infrared  and  accessible  to   the  VLTI
interferometer.   The RV  of star  B has  been monitored  in the  past
according to \citet{R10}, but these  data remain unpublished.  It will
be  easy to  obtain  a modern  88 day  spectroscopic  orbit of  Ba,Bb.
Combined with  accurate relative positions  from the VLTI,  these data
will constrain the masses in this pair.


\section{HIP 89234 (HR 6751)}
\label{sec:89324}


\begin{figure}
\epsscale{1.1}
\plotone{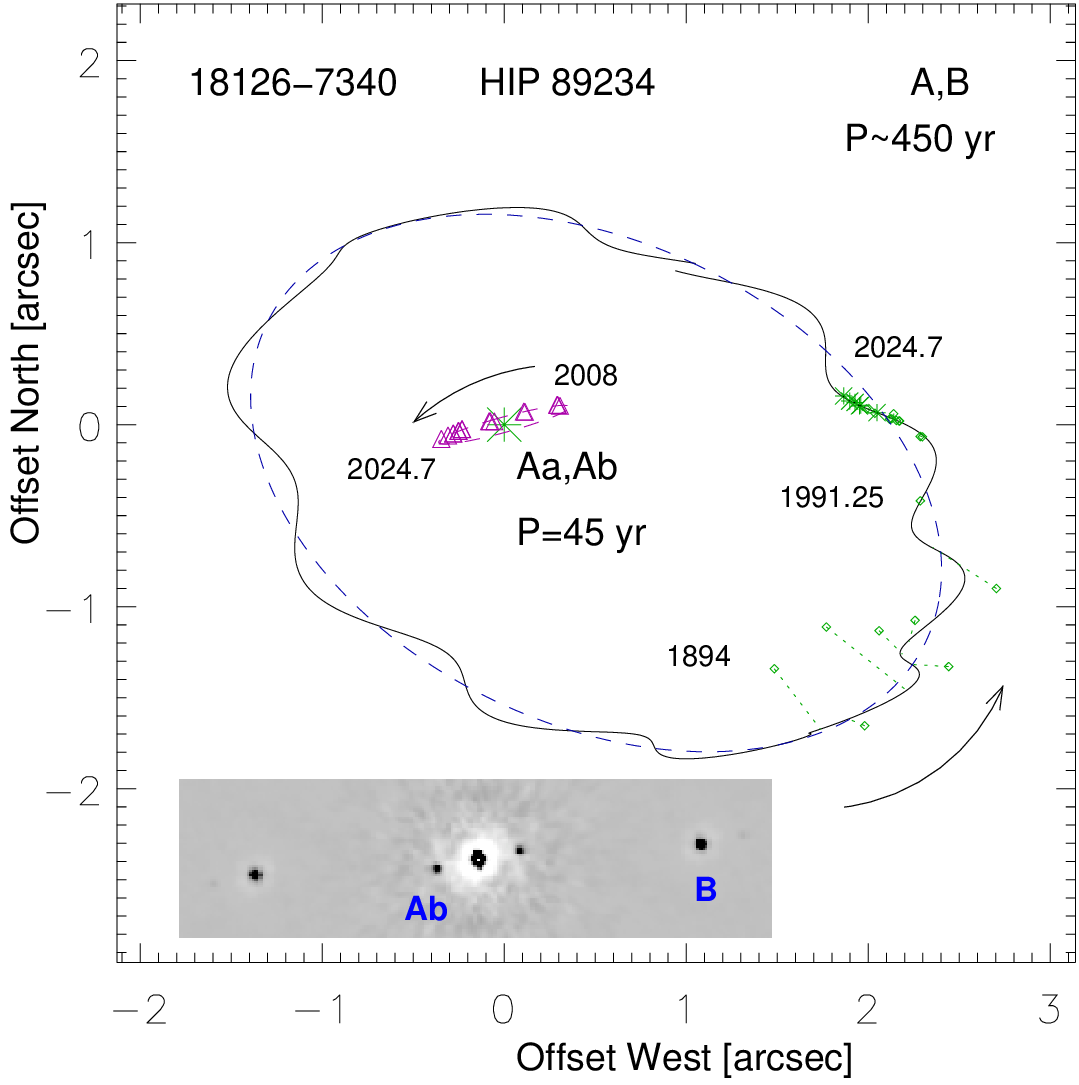}
\caption{Orbits of HIP 89234.The insert  shows the latest speckle ACF recorded
  in 2024.7 in negative rendering.
\label{fig:89234}
}
\end{figure}

This  bright ($V=5.85$  mag, F5V)  visual  triple system  is known  as
HR~6751 and  WDS J18126$-$7340.   The outer companion  B (HDO  284AB) at
2\arcsec  ~has  been  measured   visually  since  1891,  although  the
available historic  coverage of  7 micrometer measures  is sparse.
The  inner subsystem  Aa,Ab  was discovered  at SOAR  in  2008 and  is
designated as TOK~58Aa,Ab.  Both visual  companions are fainter than star A
by $\sim$3.4 mag in the $I$ band.

Gaia DR3 measured parallaxes  of 23.678$\pm$0.040 and 23.710$\pm$0.035
for  stars A  and B,  respectively, with  a modest  RUWE. During the Gaia
mission, star A moved almost linearly in the inner orbit and Ab was at
close separation,  explaining why it  has not spoiled  the astrometry.
The  ``photometric'' masses  of Aa,  Ab,  and B  estimated from  their
absolute magnitudes are 1.46, 0.73, and 0.82 \msun, respectively. Star
A   is    slightly   evolved   (above   the    main   sequence),   and
\citet{AbuArlob2023} estimated its age at 1.4\,Gyr.

\citet{Ling2016c} published a preliminary (grade  5) orbit of A,B with
a period of 324  yr. The small arc of this orbit  covered so far gives
only loose  constraints on the  elements. The inner subsystem  is also
observed over a fraction of  its orbit. Here the position measurements
are     fitted    by     two     Keplerian    orbits     simultaneously
(Figure~\ref{fig:89234}).

The measurements  of the inner pair  Aa,Ab come from SOAR,  except one
measure  in   2017.44  made  by  \citet{Horch2019}   at  the  Gemini-S
telescope,  where  the quadrant  of  Aa,Ab  had  to be  changed.   The
quadrants of all  SOAR measures are defined by the  orientation of the
outer pair  and cannot be changed.   The pair Aa,Ab was  discovered in
2008.77 at 288\degr ~position angle, went through conjunction in 2018,
and  now  opens  up  at  99\degr.    The  free  fit  gives  the  inner
eccentricity   of   0.006$\pm$0.025,   below  significance,   so   the
zero-eccentricity solution  is imposed.  The poorly  constrained outer
period is  fixed to 450\,yr  and the  outer eccentricity to  0.3 (this
combination  defines  the remaining  outer  elements  and matches  the
expected  mass  sum).  The  motion  of  B  relative  to Aa  in  2016.0
predicted by  both orbits is  $(31.1, 12.4)$  \masyr. The motion  of B
relative  to the  photocenter of  A measured  by Gaia  DR3 is  $(29.3,
12.3)$ \masyr, in  good agreement.  Given the  relatively large wobble
amplitude $f=0.37 \pm  0.02$, the speed of B relative  to A depends on
both orbits. The wobble factor  indicates the inner mass ratio $q_{\rm
  in} = 0.59$, and the photometric  masses correspond to $q_{\rm in} =
0.50$.

Using the Gaia  parallax of Aa, the  two orbits lead to  the inner and
outer mass  sums of  2.0 and 3.0  \msun, respectively.   The estimated
mass sums are  2.2 and 3.0 \msun.  Considering  the preliminary nature
of both  orbits, this agreement  is satisfactory.  I tried  to impose
different constraints on the outer orbit and found that the outer mass
sum is relatively robust,  and it can be even larger,  e.g. 3.6 \msun
if we set $e_{\rm out}=0.2$. This hints that star B may have a subsystem.

The   different  inclinations   of  the   orbits  are   quite  obvious
(Figure~\ref{fig:89234}).  The mutual  inclination  computed from  the
elements    is    either    51\degr   ~or    118\degr.  Note
also the modest  period ratio, $\sim$10.  Dynamical  stability of this
system \citep{Mardling2001} calls for $a_{\rm out}(1-e_{\rm out})/a_{\rm
  in} > 3.5$ or $e_{\rm out} < 0.35$.  Therefore, very eccentric outer
orbits are excluded, justifying the adopted value of $e_{\rm out}$.


\section{HIP 105404 (BS~Ind)}
\label{sec:105404}


\begin{figure*}
\epsscale{0.9}
\plotone{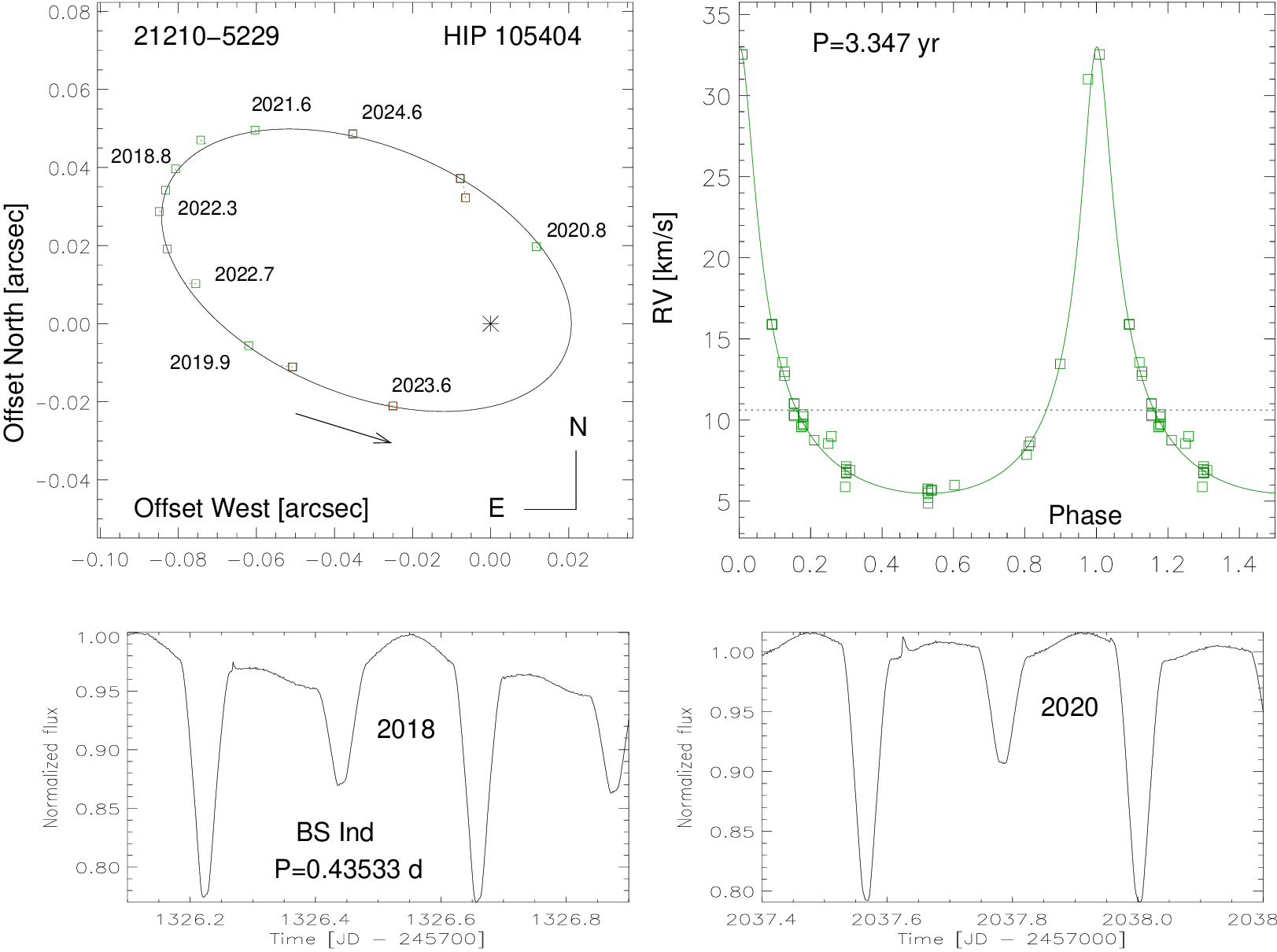}
\caption{The outer orbit (top) and the TESS light curves (bottom) of BS Ind.
\label{fig:BSInd}
}
\end{figure*}

This eclipsing triple system is designated  in the catalogs as BS Ind,
HD 202947, HIP~105404, TIC~79403459,  WDS J21210$-$5229 ($V=8.90$ mag,
K0V).   Literature suggests  its  membership  in the  Tucana-Horlogium
association with an estimated age  of $\sim$30 Myr based on kinematics
and such youth indicators as location on the color-magnitude diagrams,
presence of the lithium line,  and chromospheric activity.  The triple
nature of BS~Ind  has been discovered by  \citet{Guenther2005}. On the
one hand,  they found eclipses  with a  period of 0.435338  days using
Hipparcos photometry.  On  the other hand, the  RV variation indicated
an eccentric orbit with a period of 3.3 yr and $e=0.6$.  These authors
concluded that the RVs  refer to the K0V primary  star A of mass
0.9 \msun, while the eclipsing pair of comparable total mass is hosted
by  the  secondary   component  B.   The  RVs  of   Ba  were  measured
approximately from  the broad  feature in  the CCF,  leading to  an RV
amplitude of  65 \kms for  Ba.  The  authors suggested that  the outer
pair can be resolved interferometrically.

This star has  been observed by the  SOAR speckle camera in  2018 in a
survey of  young moving groups  and resolved into  a 90 mas  pair with
comparable components  \citep{Tokovinin2021}.  The  WDS assigned  it a
meaningless  name  CVN~66Aa,Ab.   Two  faint companions  B  and  C  at
3\farcs25  and  5\farcs86 separations,  denoted  as  CVN~66AB and  AC,
respectively,  are optical,  as  evidenced by  their Gaia  parallaxes.
Further speckle  monitoring revealed  a fast orbital  motion.  Speckle
observations accumulated  from 2018.8  to 2024.6 allow  calculation of
the combined  spectro-interferometric orbit  using the  published RVs.
New  astrometry from  Gaia  and the  photometry  from TESS  contribute
additional information on this system.

The 14  speckle measurements and  the available  RVs were fitted  by a
Keplerian orbit (Figure~\ref{fig:BSInd}).   The weighted rms residuals
are 1.4 mas in position and 0.28 \kms in RVs. The RVs were measured by
\citet{Guenther2005} in  1994.7--2004.5  and,  together with  the  recent
speckle  data,   they  accurately   constrain  the  outer   period  of
3.3474$\pm$0.0015  yr.   I  downloaded  from the  ESO  archive  three
additional FEROS  spectra taken  in 2017.545 and  computed the  RV of
15.9 \kms by cross-correlation (note the first point on the descending
branch of the RV curve).  Three RVs measured with CHIRON in 2022--2024
were used as  well.  Although the orbit plots look  good, there remain
correlations  between several  elements  (e.g.  $e$  and  $P$) in  the
least-squares fit.  A better  spectroscopic coverage of the periastron
is needed  to reduce the correlations.   Unfortunately, the periastron
in 2023.9 was missed because CHIRON was closed at that time.

Motion of the  photocenter with the 3.3 yr period  biases the parallax
and  PM   measurements  from   space  missions.   The   HIP2  parallax
\citep{HIP2}   is  22.25$\pm$1.40   mas,  the   Gaia  parallaxes   are
18.99$\pm$0.33  mas  (DR2) and  16.50$\pm$0.51  mas  (DR3).  Note  the
increased error  of the DR3  parallax caused  by the fast  motion near
periastron in  2017.  The DR2 parallax  is based on the  orbit segment
near the apastron  and appears to be less biased.   Using the mass sum
of 2.2 \msun estimated from absolute magnitudes, I adopt the dynamical
parallax of 19.0\,mas,  same as the DR2 parallax.  With  the A mass of
0.85 \msun (see below), the RV amplitude and inclination correspond to
the B mass of 1.36  \msun, matching the photometrically estimated mass
sum of Ba,Bb.

\begin{figure}
\epsscale{1.1}
\plotone{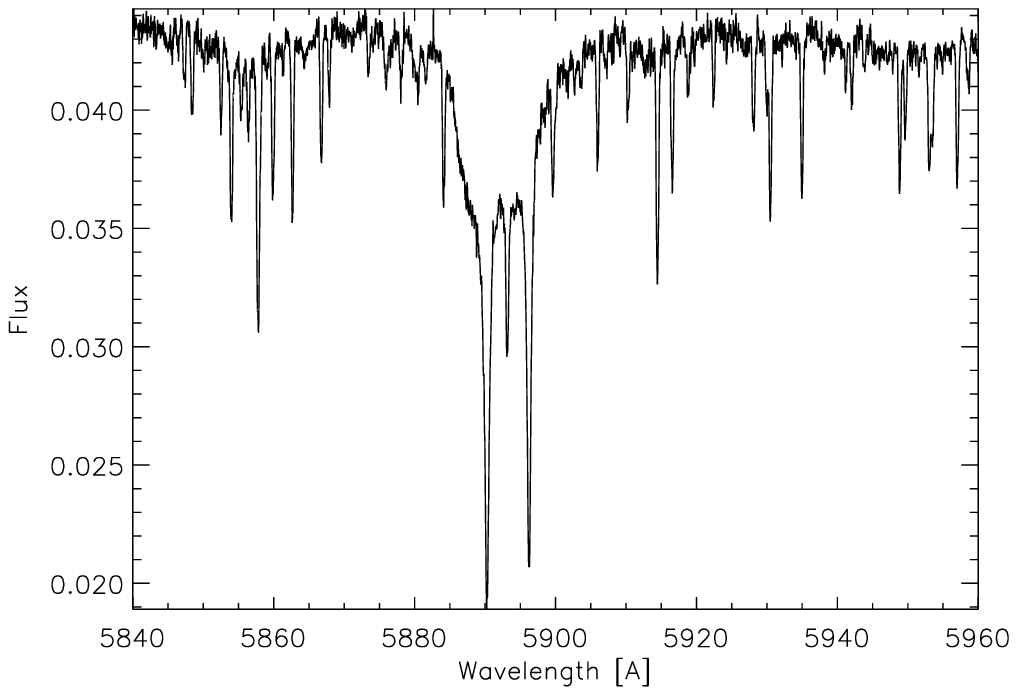}
\caption{Fragment  of the  FEROS spectrum  of BS  Ind around  sodium D
  lines taken on  2017-08-03. Note the narrow lines of  A and the wide
  and blended lines of Ba and Bb. The spectrum was downloaded from the
  ESO archive. The flux units are arbitrary.
\label{fig:Dlines}
}
\end{figure}

The  TESS light  curves (LCs)  measured in  2018 (sector  1) and  2020
(sector    27)    are    presented    in   the    lower    panels    of
Figure~\ref{fig:BSInd}. These curves are not  phase-folded.  There are a few
flares, as  expected in  these chromoshperically active  stars (flares
were   also  detected   previously   from  the   ground).   The   flux
normalization is approximate.

A striking  difference between the 2018  and 2020 LCs is  evident. The
former  has a  depression  before  the secondary  minimum  and a  hump
after. Such features are attributed usually to an asymmetric accretion
disk or  a stream around  the secondary,  although they could  also be
caused by large starspots.  The brighter  part of the disk is eclipsed
by the  primary before the  secondary minimum.  The  secondary minimum
has a small flat bottom, signaling that the eclipse is total.  Hence,
the depth of the secondary minimum,  0.10, equals the relative flux of
Bb in the TESS  band. The primary eclipse is partial,  so its depth of
$\sim$0.22 is  a lower limit on  the Ba relative flux  (I adopt below
0.30).  The Hipparcos  LC published by \citet{Guenther2005}   is noisy, but
suggests a  much smaller secondary  minimum. This can be  explained by
the lower temperature  of the secondary (less light in  the Hipparcos
band) and, possibly, a lower inclination  of the eclipsing pair due to
its  precession.  The  non-stationarity of  the LC  evidenced by  TESS
merits further investigation.

The magnitude difference between A  and B measured at SOAR establishes
the relative fluxes of A and B. It is 0.31 mag in $I$ (7 measures, rms
scatter 0.19  mag) and 0.40 mag  in $y$ (7 measures,  rms scatter 0.50
mag).   Variability of star B increases the scatter of differential
speckle photometry at SOAR.

In the following, I adopt the  flux ratio A:Ba:Bb of 0.60:0.30:0.10 in
the $T$ or  $G$ bands, in rough agreement with  the speckle photometry
 and the TESS  LCs.  The combined magnitudes recovered  from Simbad are
8.90   ($V$),  8.52   ($G$),  7.89   ($I$),  7.18   ($J$),  and   6.57
($K$). Splitting  the $V$ flux  in the indicated proportion  (which is
not quite correct, considering that B is redder than A) results in the
$V$  magnitudes of  9.46, 10.22,  and  11.41 mag  for A,  Ba, and  Bb,
respectively.  The parallax  of 19 mas and  the standard main-sequence
relation lead to  the masses of 0.86, 0.78, and  0.70 \msun.  However,
the RV amplitude  of Ba measured by \citet{Guenther2005}  imply the Bb
mass of 0.23  \msun, while the outer orbit corresponds  to the Bb mass
of 0.60 \msun (subtracting the Ba mass 0.78 from the mass of B, 1.38).
\citet{Guenther2005}  likely underestimated  the  RV  amplitude of  Ba
owing to blending.

\citet{Guenther2005} looked for  the signatures of the secondary components
in the spectrum and could not find  any. They assumed that B hosts two
similar late-M  dwarfs that are  much fainter than  A. In fact,  as we
know now,  the fluxes of A  and B are comparable  (ratio 0.6:0.4), and
the masses of Ba and Bb are  in the K-dwarf domain.  A fragment of the
FEROS    spectrum    near    the    sodium   D    lines    shown    in
Figure~\ref{fig:Dlines} makes  it clear that  the lines of Ba  and Bb
are strong, but very wide and  blended. A spectroscopic orbit of Ba,Bb
can possibly be  determined by modeling the blended  spectra.  An even
more accurate orbit of B could be derived by photometry. The estimated
outer axis of  the B orbit is about  $a \sin i \approx 0.7$  au, so a
light-time effect of $\sim$5 min is expected.

The Gaia  DR3, released in June  2022, did not provide  astrometric or
spectroscopic orbit of the outer pair,  despite its short period. As a
substitute, we  can study the  PMA measured  by Gaia DR2  (less biased
than DR3, see above).  The  long-term PM deduced by \citet{Brandt2018}
is $(36.58, -103.74)$ \masyr, in  excellent agreement with the Tycho-2
long-term PM  of $(35.4,  -101.2)$ \masyr.  Subtracting  the long-term
HIP$-$DR2 PM from  the DR2 Gaia PM yields the  PMA of $(-5.96, +7.83)$
\masyr.  The orbital motion of B  relative to A in 2015.5 was $(19.65,
-31.08)$  \masyr.    Considering  the   approximate  nature   of  this
calculation, the agreement of the  PMA direction with predictions from
the orbit  is encouraging.   The ratio of  PMA and  orbital velocities
gives the wobble factor $f^*=0.27$.  The mass ratio $q=1.35/0.86=1.57$
and the light ratio $r =  0.4/0.6=0.67$ correspond to $f^* = 0.21$, in
reasonable agreement.  Therefore, the  preliminary estimates of masses
and relative fluxes given above are supported by the Gaia astrometry.

\citet{Guenther2005} drew attention to the fact that the Ba,Bb pair is
the closest among known young  binaries.  Most eclipsing binaries with
such short periods are older than $\sim$0.6 Gyr, being produced by the
slow     angular     momentum     loss    through     stellar     wind
\citep[e.g.][]{Hwang-Zakamska2020}.  However, the young  age of BS Ind
based on its assumed  membership in the Tucana-Horlogium association
 may  be incorrect.  The long-term PM,  center-of-mass RV, and
parallax of 19.0 mas correspond to the Galactic velocity of $(U,V,W) =
(-2.1, -27.6, -9.8)$  \kms for BS Ind, while the  mean velocity of the
association   is    $(U,V,W)   =    (-10.72,   -20.2,    -0.7)$   \kms
\citep{Gagne2018}.    The   difference   largely  exceeds   both   the
measurement errors and the velocity dispersion in the association.  In
fact, \citet{Guenther2005} already noted the large discrepancy between
the mean  RVs of  BS Ind and  Tuc-Hor, but,  nevertheless, highlighted
membership in the  association in the title of their  paper.  The high
chromospheric  activity of  BS Ind  (emission lines,  X-ray radiation,
flares) could be  maintained by the fast rotation  in the synchronized
pair  Ba,Bb.   A  similar  case  of HIP  45734  containing  a  0.5 day
subsystem was  studied by \citet{Tok2020bb}: this  object, believed to
be a  pre-main sequence star,  in fact is  older.  On the  other hand,
\citet{Guenther2005} detected lithium  lines in the  spectrum, confirming
the relative youth of BS Ind.  The 2017 FEROS spectra contain a strong
and narrow Li line, but the H$\alpha$ line is in absorption, with only
a weak emission  core at the center.  The  projected rotation velocity
of A is modest, $V \sin i = 13.0 \pm 0.4$ \kms.  This triple system is
definitely not very old, but rather ``juvenile''.

The  large  inclination of  the  outer  orbit, 42\degr,  excludes  its
coplanarity  with the  inner  eclipsing  pair.  Dynamical  interaction
between the  orbits should cause precession  of the inner pair  with a
period on  the order  of $P_{\rm  out}^2/P_{\rm in}  \sim 850$  yr, or
4\degr ~in 10 yr, so a changing  depth of the LC minima can possibly be
detected.   Furthermore,  the eclipse  time  variation  caused by  the
light-time  effect   and  mutual   dynamics  can   provide  additional
constraints on this interesting triple system.

\section{Summary}
\label{sec:sum}

This work  increases the  number of  nearby hierarchical  systems with
known orbits  by a small  amount, adding 10  orbits in 6  systems. The
progress  in this  area  is slow  owing to  such  factors as  required
angular  resolution and  time coverage.   I  used the  time series  of
speckle  interferometry  measurements at  the  4.1  m SOAR  telescope,
complemented by RVs and other  data available in the literature. These
data give access to periods on the  order of a decade (the 44 yr inner
orbit is covered only partially).  Longer periods of outer systems are
determined with  a large  uncertainty using  historic data,  or simply
estimated    from   projected    separations.    Figure~\ref{fig:plps}
illustrates the  wide range of  periods addressed in this  paper, from
the shortest  0.5 day inner period  of BS~Ind to the  longest 0.5 Myr
outer period of HIP 11783.

Of  special   interest  are  two  resolved   triples  with  comparable
separations and periods, HIP 64836 (5 yr  and 30 yr) and HIP 89234 (45
yr and $\sim$450  yr). The small period ratios (6  and 10) place these
systems  near  the  limit  of dynamical  stability.   Hence  dynamical
interactions between  inner and outer  subsystems are strong  and will
eventually become measurable through  deviations from the simple model
of two  Keplerian orbits  used here.  Period  ratios close  to integer
numbers are suggestive  of the mean motion resonances,  but a complete
coverage  of outer orbits is  needed to  prove this  hypothesis.  It
looks plausible for HIP 64836,  where the orbits are near-circular and
close to  coplanarity.  This  systems joins  the class  of hierarchies
with planetary-type  architecture, presumably formed  by fragmentation
and migration  in protostellar disks \citep{Mult2021}.   Other members
of   this  class   were  discovered   and/or  studied   by  our   team
\citep{HD91962,TL2017,Dancingtwins,TL2020}.     The    large    mutual
inclination  of   orbits  in   HIP~89234  sets   it  apart   from  the
planetary-type hierarchies, although  small eccentricities match this
architecture better  than predictions for dynamically  formed triples,
where large eccentricities and misaligned orbits are expected.

Another system  of interest  is HIP  105404 (BS  Ind).  It  is compact
(outer period  3.3 yr), and  such hierarchies are  intrinsically rare.
Its   inner  eclipsing   pair  is   definitely  misaligned   with  the
well-defined outer  orbit ($i_{\rm  out} = 42\fdg3$),  suggesting that
Lidov-Kozai cycles  coupled with  tidal friction  could have  played a
role in shrinking the  inner pair. This system is not  a member of the
Tucana-Horlogium   association,  contrary   to  the   belief  of   its
discoverers \citep{Guenther2005};  signs of youth are  attributable to
the chromospheric activity of the tidally synchronized eclipsing pair.

The Gaia space  mission will provide a  time coverage of 11  yr in its
final DR5, while its spatial resolution  is limited to 0\farcs1 by the
1 m apertures. The systems studied here take advantage of the accurate
Gaia astrometry and also  reveal its limitations. Standard astrometric
models  (single  star,  binary  orbit, acceleration)  do  not  capture
complex motions in  some close triple systems.   Dedicated analysis of
individual transits (to  become public in DR4) will  be greatly helped
by  additional ground-based  data such  as speckle  interferometry and
high-resolution spectroscopy.



\begin{acknowledgments}

I am grateful to D.~Latham for providing the RVs of HD~130412 measured
at CfA and for comments on the  paper.  The research was funded by the
NSF's NOIRLab.  This  work used the SIMBAD service  operated by Centre
des   Donn\'ees   Stellaires   (Strasbourg,   France),   bibliographic
references from  the Astrophysics Data System  maintained by SAO/NASA,
and the Washington Double Star  Catalog maintained at USNO.  This work
has made use of data from the European Space Agency (ESA) mission Gaia
(\url{https:\\www.cosmos.esa.int/gaia}),  processed by  the Gaia  Data
Processing        and         Analysis        Consortium        (DPAC,
\url{https:\\www.cosmos.esa.int/web/gaia/dpac/consortium}).    Funding
for the DPAC has been provided by national institutions, in particular
the  institutions participating  in the  Gaia Multilateral  Agreement.
This research has made use of  the services of the ESO Science Archive
Facility and  data collected by  the TESS  mission funded by  the NASA
Explorer Program.

\end{acknowledgments} 

\facility{SOAR, CTIO:1.5m, Gaia, TESS}





\bibliography{triples.bib}
\bibliographystyle{aasjournal}


\end{document}